\documentclass[conference,a4paper]{APSIPA2021}
\usepackage{amsmath}
\usepackage{graphicx}
\usepackage{multirow}
\usepackage{threeparttable}
\usepackage[backend=biber,style=ieee,]{biblatex}

\usepackage{tabularx}

\usepackage{geometry}
\usepackage{fancyhdr}

\begin{document}

\title{AVATAR: Robust Voice Search Engine Leveraging Autoregressive Document Retrieval and Contrastive Learning}

\author{
\authorblockN{
Yi-Cheng Wang, Tzu-Ting Yang, Hsin-Wei Wang,  Bi-Cheng Yan, Berlin Chen
}

\authorblockA{
Department of Computer Science and Information Engineering, National Taiwan Normal University, Taiwan \\
E-mail:\{yichengwang, tzutingyang, hsinweiwang, bicheng, berlin\}@ntnu.edu.tw}
}

\maketitle
\thispagestyle{empty}
\pagestyle{empty}

\begin{abstract}
 Voice, as input, has progressively become popular on mobiles and seems to transcend almost entirely text input. Through voice, the voice search (VS) system can provide a more natural way to meet user's information needs. However, errors from the automatic speech recognition (ASR) system can be catastrophic to the VS system. Building on the recent advanced lightweight autoregressive retrieval model, which has the potential to be deployed on mobiles, leading to a more secure and personal VS assistant. This paper presents a novel study of VS leveraging autoregressive retrieval and tackles the crucial problems facing VS, viz. the performance drop caused by ASR noise, via data augmentations and contrastive learning, showing how explicit and implicit modeling the noise patterns can alleviate the problems. A series of experiments conducted on the Open-Domain Question Answering (ODSQA) confirm our approach's effectiveness and robustness in relation to some strong baseline systems.
\end{abstract}


\section{Introduction}
Compared to desktop and laptop computers, flourishing with recent advances in speech technology, voice-based input is gradually replacing text input as the primary method of interaction between humans and machines on small-screen mobile devices ~\cite{DBLP:journals/corr/abs-1902-01790}. Through voice inputs, how to provide users with an effective mechanism to access the content they want in an overwhelming amount of information available on the Internet, viz. voice search (VS), has gained a place with other spoken language technologies at the center of the stage.
VS is related to, but distinct from spoken document retrieval (SDR)~\cite{DBLP:journals/spm/LeeC05, DBLP:conf/icassp/Fan-JiangLC20}. Compared to SDR, in which text queries are given to the system and need to search for relevant spoken documents meanwhile facing essential issues, the queries are usually too short to conveying
the information needs. VS searches for relevant text documents using a more natural modality, viz. voice, that lead to longer queries, thus better-expressing information needs ~\cite{DBLP:conf/sigir/Guy16}. 

\begin{figure}[t]
  \centering
  \includegraphics[width=\linewidth]{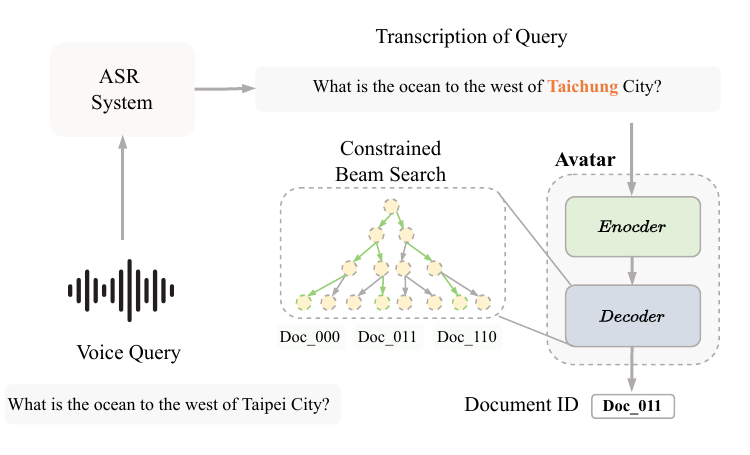}
  \caption{Flow diagram of our proposed Avatar VS system. Given a voice query, Avatar first uses an ASR system to obtain a text query (errors generated from the ASR system were denoted in orange). Then, an autoregressive retriever directly generates the ranked list of relevant document ID (docid) through constrained beam search.}
  \label{fig:main_workflow}
\end{figure}

Over the years, many efforts have been devoted to investigating deep neural network-based retrieval methods, showing good promise in many IR and SDR tasks. These models can be applied to various VS scenarios. Recently, dense retrieval models have become the primary neural-based document retrieval approach, complementary to sparse retrieval models such as TF-IDF \cite{DBLP:journals/cacm/SaltonWY75} and BM25 \cite{DBLP:journals/ipm/JonesWR00}, that matches keywords efficiently with an inverted index. Dense retrieval models ~\cite{DBLP:conf/sigir/KhattabZ20, DBLP:journals/tacl/LuanETC21}  leverage the pre-trained language models to encode information in latent semantic space, thus better capturing the semantic relationships between queries and documents. A typical dense retrieval construct of a dual encoder contains two independent neural networks optimized for embedding the queries and documents. The advantage of the dual encoder design is that the entire corpus can be encoded and indexed offline. At the inference time, the score of a query-document pair can be efficiently computed as the inner product of the corresponding query and document embeddings. 

\begin{figure*}[t]
  \centering
  \includegraphics[width=\linewidth]{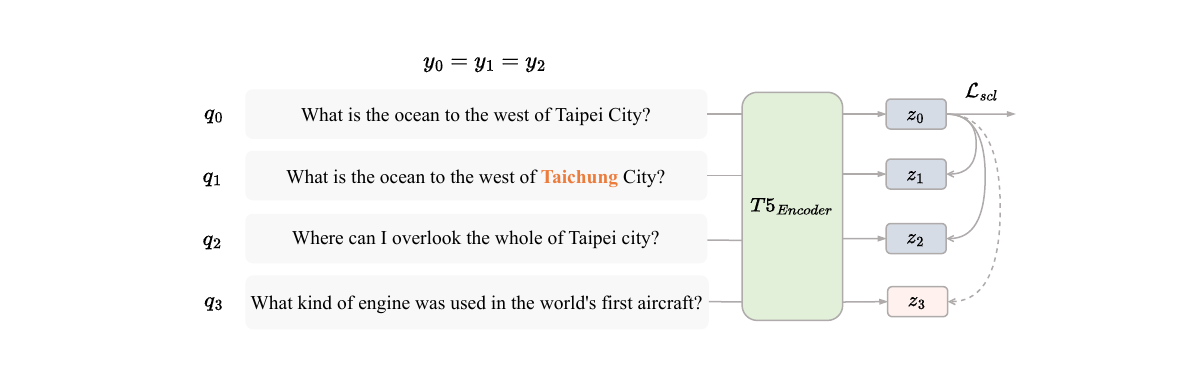}
  \caption{Pre-training: supervised contrastive learning (SCL). SCL takes data of the same class as positive samples and consists of three query types: queries from the original dataset, pseudo queries generated from the query generation model, and augmented queries.}
  \label{fig:scl_workflow}
\end{figure*}

Recently, another novel autoregressive retrieval approach orthogonal to the dual encoder models dubbed Differentiable Search Index (DSI)\cite{DBLP:journals/corr/abs-2202-06991} has been proposed.
DSI first encodes all the information about the corpus into a single transformer model's parameters space, on top of which DSI can generate the relevant document identities (docids) in an autoregressive manner in response to a user query. DSI has many advantages over the dual encoder models: 1) It avoids using only dot products, which could miss the fine-grained interaction between the query and the document meta information. 2) It lowers memory requirements; storing dense vectors for the whole corpus requires a large memory footprint. Although with the benefits mentioned above, DSI suffers serious data distribution problems during the model training and inference phase. \cite{DBLP:journals/corr/abs-2206-10128}. Specifically, DSI learns to build connections between long document texts and their docids, but at inference time, relatively short queries are input into the model to retrieve their relevant docids. Differentiable search index with query generation (DSI-QG) \cite{DBLP:journals/corr/abs-2206-10128} migrates this problem using another query generation model to generate relevant pseudo queries from documents. It uses these short pseudo queries instead of long text documents to build the connection with their docids in the training phase. After incorporating these pseudo queries \cite{DBLP:journals/corr/abs-2208-09257}, the autoregressive retrieval model has become effective while requiring less memory footprint and has potential to be deployed on the mobile or edge device, making it a more secure and personal retrieval system.

Nevertheless, as far as we are concerned, the autoregressive retrieval model has not been sufficiently and systematically studied in neither ad-hoc information retrieval (IR) nor VS, and its retrieval effectiveness is mostly unknown. Based on this background, we present an empirical VS evaluation that sheds light on the efficacy of the autoregressive retrieval model in this paper. Further, we address one of the critical problems with VS, viz. noises in the query caused by the ASR system has an enormous impact on the retrieval model \cite{DBLP:conf/cikm/SidiropoulosVK22, DBLP:conf/interspeech/ChangC22a}, by means of explicitly modeling the ASR noise pattern using data augmentation and implicitly teaching the model to distinguish the features invariant to noise using contrastive learning.
A series of experiments conducted on the Open-Domain Spoken Question Answering dataset (ODSQA) \cite{DBLP:conf/slt/LeeWCL18} confirm our approach's effectiveness and robustness in relation to some strong baseline systems.

\section{Methodology}
In this paper, we propose Avatar, a robust voice search engine leveraging \textbf{a}utoregressive retrie\textbf{va}l and con\textbf{t}rastive le\textbf{ar}ning. Figure \ref{fig:main_workflow} shows the main workflow. Given a voice query, Avatar first uses an ASR system to transcribe voice queries into a text query. Then, an autoregressive retriever built from standard transformer architecture directly generates the relevant docids through constrained beam search. In this section, we first introduce autoregressive retrieval and then demonstrate our approach to robust the model.
\subsection{Autoregressive Retrieval}
Unlike classic retrieval techniques, autoregressive retrieval methods use a sequence-to-sequence (seq2seq) language model for encoding all the corpus information into the model's parameter space. After receiving the user query, beam search is used to generate the rank list of docids. Specifically, the autoregressive retriever ranks each document $d\in D$ in the corpus by computing a score with an autoregressive formulation:
\begin{equation}
score(d|q)=p_{\theta}(y|q)=\prod_{m=1}^{M}p_{\theta}(y|y_{1<m}, q),
\label{eqn:auto_score} 
\end{equation}
where $q$ is the query entered by the user, $y$ is the set of $M$ tokens in docid of $d$, and $\theta$ represents the parameters of the model.
Original DSI models suffer from data distribution problems during the model training and inference phase. To migrate this problem, we follow the DSI-QG to use the pseudo queries generated from a seq2seq query generation model and incorporate them into the model training phase. 
In other words, we generate pseudo queries $Q_{qg}$ from their relevant documents using a query generation model. Following that we combine them with the original queries $Q_{sup}$ from the training data to form the new training sets $q\in Q_{seq}=\{Q_{sup}\cup Q_{qg}\}$. After that, we train the model by the general seq2seq training objective and teach forcing:
\begin{equation}
\mathcal{L}_{seq}(\theta)=-\sum_{q\in Q_{seq}} \log p_{\theta}(y|q).
\label{eqn:seq_loss} 
\end{equation}
A well-built docid must be able to identify the different documents while reflecting their semantic information. Since exploring different docids is not the focus of this study, we adopt the semantic docid proposed in DSI and leave the extension of docids for future work.
At first, semantic docid uses a BERT language model ~\cite{DBLP:conf/naacl/DevlinCLT19} to encode all the documents in the corpus to obtain the semantic vectors. Second, the hierarchical clustering algorithm is employed to cluster the semantically similar documents in the same group in a hierarchical fashion. Finally, we can assign each document the identifier by group number by traversing the hierarchical tree.
The docids generated by the beam search do not necessarily exist in the corpus. Inspired by \cite{DBLP:conf/iclr/CaoI0P21}, we use the constrained beam search to guide the decoder to search in a limited tokens space for each step to generate the valid docids. Concretely, we define constraints based on a prefix tree built on all docid strings.
\subsection{Explicit: Data Augmentation}
Data augmentation (DA) is a simple and effective method often used to strengthen models by explicitly exposing them to data containing ASR noise and clean data so that model learning remains invariant to noise.
Specifically, we generate three random augmentations for each query in $Q_{seq}$:
$Q_{da}=Aug(Q_{seq})$, where $Aug(\cdot)$ is a data augmentation module similar to \cite{DBLP:conf/nips/LengTZXLLQLLL21}, we generate random substitution, deletion, or insertion errors. For substitution, we use similar phonological words to substitute them. Finally, we train the model using the training sets consisting of $Q=\{Q_{seq}\cup Q_{da}\}$.

\subsection{Implicit: Contrastive Learning}
Contrastive learning (CL) can help the model distinguish the invariant features in the ASR noise and clean text query. Firstly, we pre-train the model's encoder to bring the original queries closer to the augmented queries and push the others away. By closely looking into the autoregressive retrieval model, we find that it's fundamentally similar to the sequence classification task. Thus bringing the same class close together in the latent space can benefit the classification task later. 

Supervised Contrastive Learning (SCL)\cite{DBLP:conf/nips/KhoslaTWSTIMLK20} takes data of the same class as positive samples and pulls their embeddings closer together. In the end, representations from the same class form a clustering effect and discriminate different classes by the margins created between them. It is more suitable to use SCL for the autoregressive retrieval model to enhance its robustness against ASR errors. 
Given a min-Batch of $N$ queries, random sample from the $Q$ training sets, $B=\{(q_{i}, y_{i})\}_{i=1..N}$. We first obtain the representation of each query through the model's encoder $Enc(\cdot)$ and use the $Proj(\cdot)$ projection network, $z=Proj(Enc(q))$ to obtain the sequence-level representation. Let $i\in I=\{1,... ,N\}$ be the index of $B$. The following equation can describe our SCL objective:
\begin{equation}
\mathcal L_{scl}=-\sum_{i\in I}^{}\frac{1}{|S(i)|} \sum_{s\in S(i)}^{}\log\frac{\exp(z_{i}\cdot z_{s})}{\sum_{a \in A(i)}^{}\exp(z_{i}\cdot z_{a})}
\label{eqn:scl_loss} 
\end{equation}

Here $\cdot$ denotes inner product operations, $A(i)=I\setminus\lbrace i \rbrace$ is the set of index $I$ minus $i$, $S(i)=\lbrace s\in A(i) : y_{s}=y_{i} \rbrace$ is the index of all positive samples of index $i$, and $|S( i)|$ is its cardinality. The process is illustrated in Figure \ref{fig:scl_workflow}.

\subsection{Combine}
Combining the above Explicit and Implicit approaches makes our proposed Avatar model more robust when encountering ASR noises. Overall, we first use SCL Eq. (\ref{eqn:scl_loss}) to pre-train the model's encoder and then use DA with general seq2seq objective Eq. (\ref{eqn:seq_loss}) to fine-tune the model further.

\begin{table}[]
\caption{Data statistics of ODSQA-test. We employ three different ASR systems to understand better the influences caused by different WER-level and Entity Error Rates (EER).}
\label{tab:wer_eer}
\resizebox{\columnwidth}{!}{
\begin{tabular}{lc|cc}
\hline
 & \multicolumn{1}{l|}{} & \multicolumn{2}{c}{ODSQA(Test)} \\
\multicolumn{1}{c}{\textbf{ASR Model}} & \textbf{Train Sets} & \textbf{WER} & \textbf{EER} \\ \hline
iFLYTEK & - & 10.61\% & 19.75\% \\
Conformer Mask-CTC & In-house & 15.06\% & 42.68\% \\
Conformer & Aishell & 23.89\% & 50.87\% \\ \hline
\end{tabular}
}
\end{table}

\begin{table*}[]
\centering
\label{main-result-table}
\caption{The main retrieval results of Avatar and some well-known retrieval models.}
\begin{tabular}{c|cccccccc}
\hline
\multicolumn{1}{l|}{} & \multicolumn{8}{c}{ODSQA(Test)} \\
\textbf{} & \multicolumn{2}{c}{Clean} & \multicolumn{2}{c}{WER 10\%} & \multicolumn{2}{c}{WER 15\%} & \multicolumn{2}{c}{WER 23\%} \\
\textbf{Model} & \textbf{Hit@1} & \textbf{Hit@10} & \textbf{Hit@1} & \textbf{Hit@10} & \textbf{Hit@1} & \textbf{Hit@10} & \textbf{Hit@1} & \textbf{Hit@10} \\ \hline
BM25 & 34.90 & 53.55 & 29.82 & 45.25 & 22.62 & 37.18 & 19.95 & 30.69 \\
DPR & 49.11 & \textbf{70.15} & 39.59 & 59.59 & 29.46 & 49.96 & 22.76 & 41.55 \\ \hline
DSI & 43.16 & 60.17 & 32.71 & 49.04 & 25.76 & 42.03 & 19.27 & 33.01 \\
DSI-QG & 46.51 & 64.95 & 37.22 & 54.64 & 28.43 & 44.77 & 22.76 & 38.07 \\
Avatar & \textbf{52.73} & 69.12 & \textbf{43.23} & \textbf{60.17} & \textbf{36.02} & \textbf{53.86} & \textbf{29.18} & \textbf{44.90} \\ \hline
\end{tabular}
\end{table*}

\section{Experiments}
\subsection{Experimental Setting}
\textbf{Dataset and Evaluation.}
We used Open-Domain Spoken Question Answering (ODSQA) for our experiments. ODSQA consists of 30,461 query-document pairs, whose the queries are natural language, and the 2,051 documents are from Delta Reading Comprehension Dataset (DRCD) \cite{DBLP:journals/corr/abs-1806-00920}. Since ODSQA only releases 1,465 query audios equipped with official ASR transcriptions from the ODSQA-test, we adopt these as our testing set and the remaining 28,996 query-document pairs as our training set. To observe the influences on the model caused by different WER-level and Entity Error Rates (EER), in addition to official ODSQA ASR transcriptions, we apply another two ASR systems to obtain the testing set's transcriptions. Specifically, we used a Conformer Mask-CTC ASR system trained on In-house Data and a Conformer ASR system on Aishell \cite{DBLP:conf/ococosda/BuDNWZ17}. Detail summarized in Table \ref{tab:wer_eer}.

Like the original DSI, we utilize Hits@1 and Hits@10 to evaluate the effectiveness of the baselines and the model. This metric reports the proportion of the correct docid ranked in the top 1 and top 10 predictions.
\\
\\
\textbf{Baselines.}
We compare Avatar with the following baselines:
1) Okapi BM25: A classic sparse retrieval method bases on the inverted index. 2) DPR: A dual encoder dense retriever that trained with contrastive loss and hard negatives. 3) DSI: An autoregressive retrieval method that uses document texts as input for indexing. 4) DSI-QG: An improved version of DSI that mitigates the data distribution mismatch problems by using generated queries as inputs for indexing.
\\
\\
\textbf{Implementation details}
In this study, we use two multi-lingual T5(mT5)\cite{DBLP:conf/naacl/XueCRKASBR21} base models provided by huggingface for our system, one as the Avatar model and the other as the Query Generation model. We apply the same training method as that of DSI-QG for the query generation model and generate three relevant pseudo queries for each document. For the Avatar model pre-training, we employ the same $MeanPooling$ as that of \cite{DBLP:conf/acl/NiACMHCY22} to obtain the sequence-level represents. We utilize the same mT5 model as a fair comparison for all the learning-based models in the baseline. Since the document length in the corpus is too long for the model to accommodate, we only keep the document title and the first 100 tokens as the model's inputs.

\begin{table}[]
\centering
\begingroup
\setlength{\tabcolsep}{3pt} 
\renewcommand{\arraystretch}{1} 
\caption{Ablation study of different WER-level.}
\label{tab:ablation_wer}
\begin{tabular}{c|cccccc}
\hline
\multicolumn{1}{l|}{} & \multicolumn{6}{c}{ODSQA(Test)} \\
\textbf{} & \multicolumn{2}{c}{WER 10\%} & \multicolumn{2}{c}{WER 15\%} & \multicolumn{2}{c}{WER 23\%} \\
\textbf{Model} & \textbf{Hit@1} & \textbf{Hit@10} & \textbf{Hit@1} & \textbf{Hit@10} & \textbf{Hit@1} & \textbf{Hit@10} \\ \hline
w/o Data Augm. & 37.22 & 54.64 & 28.43 & 44.77 & 22.76 & 38.07 \\
w/o SCL & 39.75 & 58.33 & 31.85 & 50.85 & 26.17 & 42.31 \\
Avatar & \textbf{43.23} & \textbf{60.17} & \textbf{36.02} & \textbf{53.86} & \textbf{29.18} & \textbf{44.90} \\ \hline
\end{tabular}
\endgroup
\end{table}

\subsection{Main Results}
The evaluation performance is presented in Table 3. Based on term matching, we find that the BM25 model has the lowest performance among all models, showing the importance of assessing the semantic information. By looking at DSI-QG, incorporating the pseudo queries can effectively improve the model's retrieval ability compared to the autoregressive DSI model. Moreover, the noise caused by the ASR system can cause catastrophic impacts on all the baselines model, including the strong DPR model. The magnitude of the drop also rises with the increase of WER. Through data augmentation and supervised contrastive learning, our proposed Avatar can alleviate the influences from the imperfection of the ASR transcriptions and not only keep the performance in a clean environment but also simultaneously increase the retrieval ability in contrast to other strong baseline systems.

\subsection{Ablation Study}
\textbf{Study of different WER.}
We now focus on evaluating the effectiveness of different ASR noise-robust methods in the model and show the performance in Table \ref{tab:ablation_wer}. Not surprisingly, explicit modeling of the noise pattern, viz. data augmentation, has the biggest effect on the model's performance. Through implicitly teaching the model to distinguish the invariant features between the ASR noise and clean query text, viz. pre-trained with SCL objective, we can further enhance the model's robustness, after combine these two, the model's performance increases vastly.
\\
\\
\\
\\

\textbf{Study of different EER.}
The entities mentioned in the query significantly impact the success of retrieving the relevant documents. 
To further test our proposed Avatar's robustness, we split the testing sets into two subsets, one contains the ASR noise in the entity mentioned in the queries, and the other includes the noise in the non-entity position, results shown in Table 4. First, the model's retrieval effect is dramatically reduced when ASR noise occurs in Entity queries, which illustrates the importance of the entities mentioned in the query for retrieval. We find that adding data augmentation and contrastive learning improves the effectiveness of our model.

\section{Conclusions and Future Work}
\label{sec:majhead}
In this paper, we have presented a novel robustness method to improve the performance of the autoregressive retrieval model when exposed to noisy ASR transcriptions, whose effectiveness has also been validated and analyzed through a series of empirical experiments. The proposed model Avatar sheds light on the potential of an on-device VS engine, which can bring more convenience, security, and personal experience. In future work, we intend to first tackle the more intensive tasks which are entities error caused by ASR system and second extend the model to a large scale that requires designing the autoregressive retrieval with more capacity. 

\begin{table}[]
\centering
\label{tab:ablation_eer}
\caption{Ablation study of different EER-level.}
\begin{tabular}{c|cccc}
\hline
\multicolumn{1}{l|}{} & \multicolumn{4}{c}{ODSQA(Test) - WER 10\%} \\
\textbf{} & \multicolumn{2}{c}{Non-entity Utterances} & \multicolumn{2}{c}{Entity Utterances} \\
\textbf{Model} & \textbf{Hit@1} & \textbf{Hit@10} & \textbf{Hit@1} & \textbf{Hit@10} \\ \hline
w/o Data Augm. & 41.16 & 58.37 & 27.86 & 41.25 \\
w/o SCL & 43.98 & 62.38 & 27.04 & 46.17 \\
Avatar & \textbf{46.90} & \textbf{64.38} & \textbf{32.24} & \textbf{47.54} \\ \hline
\end{tabular}%
\end{table}

\newpage










\printbibliography

@article{DBLP:journals/corr/abs-1902-01790,
  author    = {Fabio Crestani and
               Stefano Mizzaro and
               Ivan Scagnetto},
  title     = {Mobile Information Retrieval},
  journal   = {CoRR},
  volume    = {abs/1902.01790},
  year      = {2019},
%   url       = {http://arxiv.org/abs/1902.01790},
  eprinttype = {arXiv},
%   eprint    = {1902.01790},
  timestamp = {Tue, 21 May 2019 18:03:38 +0200},
%   biburl    = {https://dblp.org/rec/journals/corr/abs-1902-01790.bib},
%   bibsource = {dblp computer science bibliography, https://dblp.org}
}

@article{DBLP:journals/spm/LeeC05,
  author    = {Lin{-}Shan Lee and
               Berlin Chen},
  title     = {Spoken document understanding and organization},
  journal   = {{IEEE} Signal Process. Mag.},
  volume    = {22},
  number    = {5},
  pages     = {42--60},
  year      = {2005},
%   url       = {https://doi.org/10.1109/MSP.2005.1511823},
  doi       = {10.1109/MSP.2005.1511823},
  timestamp = {Wed, 24 Feb 2021 15:26:54 +0100},
%   biburl    = {https://dblp.org/rec/journals/spm/LeeC05.bib},
%   bibsource = {dblp computer science bibliography, https://dblp.org}
}

@inproceedings{DBLP:conf/icassp/Fan-JiangLC20,
  author    = {Shao{-}Wei Fan{-}Jiang and
               Tien{-}Hong Lo and
               Berlin Chen},
  title     = {Spoken Document Retrieval Leveraging Bert-Based Modeling and Query
               Reformulation},
%   booktitle = {2020 {IEEE} International Conference on Acoustics, Speech and Signal
%               Processing, {ICASSP} 2020, Barcelona, Spain, May 4-8, 2020},
  pages     = {8144--8148},
  publisher = {{IEEE}},
  year      = {2020},
%   url       = {https://doi.org/10.1109/ICASSP40776.2020.9052910},
  doi       = {10.1109/ICASSP40776.2020.9052910},
  timestamp = {Thu, 23 Jul 2020 16:20:10 +0200},
%   biburl    = {https://dblp.org/rec/conf/icassp/Fan-JiangLC20.bib},
%   bibsource = {dblp computer science bibliography, https://dblp.org}
}

@inproceedings{DBLP:conf/sigir/Guy16,
  author    = {Ido Guy},
  editor    = {Raffaele Perego and
               Fabrizio Sebastiani and
               Javed A. Aslam and
               Ian Ruthven and
               Justin Zobel},
  title     = {Searching by Talking: Analysis of Voice Queries on Mobile Web Search},
  pages     = {35--44},
  publisher = {{ACM}},
  year      = {2016},
%   url       = {https://doi.org/10.1145/2911451.2911525},
  doi       = {10.1145/2911451.2911525},
  timestamp = {Wed, 14 Nov 2018 10:58:11 +0100},
%   biburl    = {https://dblp.org/rec/conf/sigir/Guy16.bib},
%   bibsource = {dblp computer science bibliography, https://dblp.org}
}

@article{DBLP:journals/cacm/SaltonWY75,
  author    = {Gerard Salton and
               Anita Wong and
               Chung{-}Shu Yang},
  title     = {A Vector Space Model for Automatic Indexing},
  journal   = {Commun. {ACM}},
  volume    = {18},
  number    = {11},
  pages     = {613--620},
  year      = {1975},
%   url       = {https://doi.org/10.1145/361219.361220},
  doi       = {10.1145/361219.361220},
  timestamp = {Thu, 20 May 2021 08:27:15 +0200},
%   biburl    = {https://dblp.org/rec/journals/cacm/SaltonWY75.bib},
%   bibsource = {dblp computer science bibliography, https://dblp.org}
}

@article{DBLP:journals/ipm/JonesWR00,
  author    = {Karen Sparck Jones and
               Steve Walker and
               Stephen E. Robertson},
  title     = {A probabilistic model of information retrieval: development and comparative
               experiments - Part 1},
  journal   = {Inf. Process. Manag.},
  volume    = {36},
  number    = {6},
  pages     = {779--808},
  year      = {2000},
%   url       = {https://doi.org/10.1016/S0306-4573(00)00015-7},
  doi       = {10.1016/S0306-4573(00)00015-7},
  timestamp = {Thu, 14 Oct 2021 08:59:18 +0200},
%   biburl    = {https://dblp.org/rec/journals/ipm/JonesWR00.bib},
%   bibsource = {dblp computer science bibliography, https://dblp.org}
}

@inproceedings{DBLP:conf/sigir/KhattabZ20,
  author    = {Omar Khattab and
               Matei Zaharia},
  editor    = {Jimmy X. Huang and
               Yi Chang and
               Xueqi Cheng and
               Jaap Kamps and
               Vanessa Murdock and
               Ji{-}Rong Wen and
               Yiqun Liu},
  title     = {ColBERT: Efficient and Effective Passage Search via Contextualized
               Late Interaction over {BERT}},
%   booktitle = {Proceedings of the 43rd International {ACM} {SIGIR} conference on
%               research and development in Information Retrieval, {SIGIR} 2020, Virtual
%               Event, China, July 25-30, 2020},
  pages     = {39--48},
  publisher = {{ACM}},
  year      = {2020},
%   url       = {https://doi.org/10.1145/3397271.3401075},
  doi       = {10.1145/3397271.3401075},
  timestamp = {Sun, 02 Oct 2022 16:15:14 +0200},
%   biburl    = {https://dblp.org/rec/conf/sigir/KhattabZ20.bib},
%   bibsource = {dblp computer science bibliography, https://dblp.org}
}

@article{DBLP:journals/tacl/LuanETC21,
  author    = {Yi Luan and
               Jacob Eisenstein and
               Kristina Toutanova and
               Michael Collins},
  title     = {Sparse, Dense, and Attentional Representations for Text Retrieval},
  journal   = {Trans. Assoc. Comput. Linguistics},
  volume    = {9},
  pages     = {329--345},
  year      = {2021},
%   url       = {https://doi.org/10.1162/tacl\_a\_00369},
  doi       = {10.1162/tacl\_a\_00369},
  timestamp = {Fri, 10 Jun 2022 10:35:17 +0200},
%   biburl    = {https://dblp.org/rec/journals/tacl/LuanETC21.bib},
%   bibsource = {dblp computer science bibliography, https://dblp.org}
}

@article{DBLP:journals/corr/abs-2202-06991,
  author    = {Yi Tay and
               Vinh Q. Tran and
               Mostafa Dehghani and
               Jianmo Ni and
               Dara Bahri and
               Harsh Mehta and
               Zhen Qin and
               Kai Hui and
               Zhe Zhao and
               Jai Prakash Gupta and
               Tal Schuster and
               William W. Cohen and
               Donald Metzler},
  title     = {Transformer Memory as a Differentiable Search Index},
  journal   = {CoRR},
  volume    = {abs/2202.06991},
  year      = {2022},
%   url       = {https://arxiv.org/abs/2202.06991},
  eprinttype = {arXiv},
  eprint    = {2202.06991},
  timestamp = {Tue, 15 Mar 2022 13:23:14 +0100},
%   biburl    = {https://dblp.org/rec/journals/corr/abs-2202-06991.bib},
%   bibsource = {dblp computer science bibliography, https://dblp.org}
}

@article{DBLP:journals/corr/abs-2206-10128,
  author    = {Shengyao Zhuang and
               Houxing Ren and
               Linjun Shou and
               Jian Pei and
               Ming Gong and
               Guido Zuccon and
               Daxin Jiang},
  title     = {Bridging the Gap Between Indexing and Retrieval for Differentiable
               Search Index with Query Generation},
  journal   = {CoRR},
  volume    = {abs/2206.10128},
  year      = {2022},
%   url       = {https://doi.org/10.48550/arXiv.2206.10128},
  doi       = {10.48550/arXiv.2206.10128},
  eprinttype = {arXiv},
  eprint    = {2206.10128},
  timestamp = {Mon, 27 Jun 2022 16:51:57 +0200},
%   biburl    = {https://dblp.org/rec/journals/corr/abs-2206-10128.bib},
%   bibsource = {dblp computer science bibliography, https://dblp.org}
}

@article{DBLP:journals/corr/abs-2208-09257,
  author    = {Yujia Zhou and
               Jing Yao and
               Zhicheng Dou and
               Ledell Wu and
               Peitian Zhang and
               Ji{-}Rong Wen},
  title     = {Ultron: An Ultimate Retriever on Corpus with a Model-based Indexer},
  journal   = {CoRR},
  volume    = {abs/2208.09257},
  year      = {2022},
%   url       = {https://doi.org/10.48550/arXiv.2208.09257},
  doi       = {10.48550/arXiv.2208.09257},
  eprinttype = {arXiv},
  eprint    = {2208.09257},
  timestamp = {Mon, 22 Aug 2022 15:19:07 +0200},
%   biburl    = {https://dblp.org/rec/journals/corr/abs-2208-09257.bib},
%   bibsource = {dblp computer science bibliography, https://dblp.org}
}

@inproceedings{DBLP:conf/slt/LeeWCL18,
  author    = {Chia{-}Hsuan Lee and
               Shang{-}Ming Wang and
               Huan{-}Cheng Chang and
               Hung{-}yi Lee},
  title     = {{ODSQA:} Open-Domain Spoken Question Answering Dataset},
  booktitle = {2018 {IEEE} Spoken Language Technology Workshop, {SLT} 2018, Athens,
               Greece, December 18-21, 2018},
  pages     = {949--956},
  publisher = {{IEEE}},
  year      = {2018},
%   url       = {https://doi.org/10.1109/SLT.2018.8639505},
  doi       = {10.1109/SLT.2018.8639505},
  timestamp = {Wed, 03 Nov 2021 08:48:34 +0100},
%   biburl    = {https://dblp.org/rec/conf/slt/LeeWCL18.bib},
%   bibsource = {dblp computer science bibliography, https://dblp.org}
}

@inproceedings{DBLP:conf/interspeech/ChangC22a,
  author    = {Ya{-}Hsin Chang and
               Yun{-}Nung Chen},
  editor    = {Hanseok Ko and
               John H. L. Hansen},
  title     = {Contrastive Learning for Improving {ASR} Robustness in Spoken Language
               Understanding},
  booktitle = {Interspeech 2022, 23rd Annual Conference of the International Speech
               Communication Association, Incheon, Korea, 18-22 September 2022},
  pages     = {3458--3462},
  publisher = {{ISCA}},
  year      = {2022},
%   url       = {https://doi.org/10.21437/Interspeech.2022-781},
  doi       = {10.21437/Interspeech.2022-781},
  timestamp = {Wed, 12 Oct 2022 10:48:55 +0200},
%   biburl    = {https://dblp.org/rec/conf/interspeech/ChangC22a.bib},
%   bibsource = {dblp computer science bibliography, https://dblp.org}
}

@inproceedings{DBLP:conf/cikm/SidiropoulosVK22,
  author    = {Georgios Sidiropoulos and
               Svitlana Vakulenko and
               Evangelos Kanoulas},
  editor    = {Mohammad Al Hasan and
               Li Xiong},
  title     = {On the Impact of Speech Recognition Errors in Passage Retrieval for
               Spoken Question Answering},
  booktitle = {Proceedings of the 31st {ACM} International Conference on Information
               {\&} Knowledge Management, Atlanta, GA, USA, October 17-21, 2022},
  pages     = {4485--4489},
  publisher = {{ACM}},
  year      = {2022},
%   url       = {https://doi.org/10.1145/3511808.3557662},
  doi       = {10.1145/3511808.3557662},
  timestamp = {Wed, 19 Oct 2022 17:09:02 +0200},
%   biburl    = {https://dblp.org/rec/conf/cikm/SidiropoulosVK22.bib},
%   bibsource = {dblp computer science bibliography, https://dblp.org}
}

@inproceedings{DBLP:conf/nips/LengTZXLLQLLL21,
  author    = {Yichong Leng and
               Xu Tan and
               Linchen Zhu and
               Jin Xu and
               Renqian Luo and
               Linquan Liu and
               Tao Qin and
               Xiangyang Li and
               Edward Lin and
               Tie{-}Yan Liu},
  editor    = {Marc'Aurelio Ranzato and
               Alina Beygelzimer and
               Yann N. Dauphin and
               Percy Liang and
               Jennifer Wortman Vaughan},
  title     = {FastCorrect: Fast Error Correction with Edit Alignment for Automatic
               Speech Recognition},
  pages     = {21708--21719},
  year      = {2021},
%   url       = {https://proceedings.neurips.cc/paper/2021/hash/b597460c506e8e35fb0cc1c1905dd3bc-Abstract.html},
  timestamp = {Tue, 03 May 2022 16:20:49 +0200},
%   biburl    = {https://dblp.org/rec/conf/nips/LengTZXLLQLLL21.bib},
%   bibsource = {dblp computer science bibliography, https://dblp.org}
}

@inproceedings{DBLP:conf/iclr/CaoI0P21,
  author    = {Nicola De Cao and
               Gautier Izacard and
               Sebastian Riedel and
               Fabio Petroni},
  title     = {Autoregressive Entity Retrieval},
  booktitle = {9th International Conference on Learning Representations, {ICLR} 2021,
               Virtual Event, Austria, May 3-7, 2021},
  publisher = {OpenReview.net},
  year      = {2021},
%   url       = {https://openreview.net/forum?id=5k8F6UU39V},
  timestamp = {Wed, 23 Jun 2021 17:36:39 +0200},
%   biburl    = {https://dblp.org/rec/conf/iclr/CaoI0P21.bib},
%   bibsource = {dblp computer science bibliography, https://dblp.org}
}

@article{DBLP:journals/corr/abs-1806-00920,
  author    = {Chih{-}Chieh Shao and
               Trois Liu and
               Yuting Lai and
               Yiying Tseng and
               Sam Tsai},
  title     = {{DRCD:} a Chinese Machine Reading Comprehension Dataset},
  journal   = {CoRR},
  volume    = {abs/1806.00920},
  year      = {2018},
%   url       = {http://arxiv.org/abs/1806.00920},
  eprinttype = {arXiv},
  eprint    = {1806.00920},
  timestamp = {Mon, 13 Aug 2018 16:48:22 +0200},
%   biburl    = {https://dblp.org/rec/journals/corr/abs-1806-00920.bib},
%   bibsource = {dblp computer science bibliography, https://dblp.org}
}

@inproceedings{DBLP:conf/ococosda/BuDNWZ17,
  author    = {Hui Bu and
               Jiayu Du and
               Xingyu Na and
               Bengu Wu and
               Hao Zheng},
  title     = {{AISHELL-1:} An open-source Mandarin speech corpus and a speech recognition
               baseline},
  pages     = {1--5},
  publisher = {{IEEE}},
  year      = {2017},
%   url       = {https://doi.org/10.1109/ICSDA.2017.8384449},
  doi       = {10.1109/ICSDA.2017.8384449},
  timestamp = {Wed, 16 Oct 2019 14:14:56 +0200},
%   biburl    = {https://dblp.org/rec/conf/ococosda/BuDNWZ17.bib},
%   bibsource = {dblp computer science bibliography, https://dblp.org}
}

@inproceedings{DBLP:conf/acl/NiACMHCY22,
  author    = {Jianmo Ni and
               Gustavo Hernandez {\'{A}}brego and
               Noah Constant and
               Ji Ma and
               Keith B. Hall and
               Daniel Cer and
               Yinfei Yang},
  editor    = {Smaranda Muresan and
               Preslav Nakov and
               Aline Villavicencio},
  title     = {Sentence-T5: Scalable Sentence Encoders from Pre-trained Text-to-Text
               Models},
  booktitle = {Findings of the Association for Computational Linguistics: {ACL} 2022,
              Dublin, Ireland, May 22-27, 2022},
  pages     = {1864--1874},
  publisher = {Association for Computational Linguistics},
  year      = {2022},
%   url       = {https://doi.org/10.18653/v1/2022.findings-acl.146},
  doi       = {10.18653/v1/2022.findings-acl.146},
  timestamp = {Mon, 01 Aug 2022 16:27:44 +0200},
%   biburl    = {https://dblp.org/rec/conf/acl/NiACMHCY22.bib},
%   bibsource = {dblp computer science bibliography, https://dblp.org}
}

@inproceedings{DBLP:conf/naacl/DevlinCLT19,
  author    = {Jacob Devlin and
               Ming{-}Wei Chang and
               Kenton Lee and
               Kristina Toutanova},
  editor    = {Jill Burstein and
               Christy Doran and
               Thamar Solorio},
  title     = {{BERT:} Pre-training of Deep Bidirectional Transformers for Language
               Understanding},
  pages     = {4171--4186},
  publisher = {Association for Computational Linguistics},
  year      = {2019},
%   url       = {https://doi.org/10.18653/v1/n19-1423},
  doi       = {10.18653/v1/n19-1423},
  timestamp = {Mon, 26 Sep 2022 12:21:55 +0200},
%   biburl    = {https://dblp.org/rec/conf/naacl/DevlinCLT19.bib},
%   bibsource = {dblp computer science bibliography, https://dblp.org}
}

@inproceedings{DBLP:conf/nips/KhoslaTWSTIMLK20,
  author    = {Prannay Khosla and
               Piotr Teterwak and
               Chen Wang and
               Aaron Sarna and
               Yonglong Tian and
               Phillip Isola and
               Aaron Maschinot and
               Ce Liu and
               Dilip Krishnan},
  editor    = {Hugo Larochelle and
               Marc'Aurelio Ranzato and
               Raia Hadsell and
               Maria{-}Florina Balcan and
               Hsuan{-}Tien Lin},
  title     = {Supervised Contrastive Learning},
  booktitle = {Advances in Neural Information Processing Systems 33: Annual Conference
              on Neural Information Processing Systems 2020, NeurIPS 2020, December
              6-12, 2020, virtual},
  year      = {2020},
%   url       = {https://proceedings.neurips.cc/paper/2020/hash/d89a66c7c80a29b1bdbab0f2a1a94af8-Abstract.html},
%   timestamp = {Tue, 19 Jan 2021 15:57:20 +0100},
%   biburl    = {https://dblp.org/rec/conf/nips/KhoslaTWSTIMLK20.bib},
%   bibsource = {dblp computer science bibliography, https://dblp.org}
}

@inproceedings{DBLP:conf/naacl/XueCRKASBR21,
  author    = {Linting Xue and
               Noah Constant and
               Adam Roberts and
               Mihir Kale and
               Rami Al{-}Rfou and
               Aditya Siddhant and
               Aditya Barua and
               Colin Raffel},
  editor    = {Kristina Toutanova and
               Anna Rumshisky and
               Luke Zettlemoyer and
               Dilek Hakkani{-}T{\"{u}}r and
               Iz Beltagy and
               Steven Bethard and
               Ryan Cotterell and
               Tanmoy Chakraborty and
               Yichao Zhou},
  title     = {mT5: {A} Massively Multilingual Pre-trained Text-to-Text Transformer},
  booktitle = {Proceedings of the 2021 Conference of the North American Chapter of
              the Association for Computational Linguistics: Human Language Technologies,
              {NAACL-HLT} 2021, Online, June 6-11, 2021},
  pages     = {483--498},
  publisher = {Association for Computational Linguistics},
  year      = {2021}
%   biburl    = {https://dblp.org/rec/conf/naacl/XueCRKASBR21.bib},
%   bibsource = {dblp computer science bibliography, https://dblp.org}
}

\end{document}